\documentclass{article}

\newcommand{\chemfig}[1]{\ensuremath{\mathrm{#1}}}

\usepackage{epstopdf}
\usepackage{url}

\usepackage{todonotes}
\begin{document}

\title{An Intermediate Level of Abstraction for Computational Systems Chemistry}
\author{
\\
Jakob~L.~Andersen$^{1,2}$,
Christoph Flamm$^{3,4}$,\\
Daniel Merkle$^{1}$\footnote{Email: daniel@imada.sdu.dk},
Peter~F.~Stadler$^{4-9}$
\\[1cm]
 $^{1}$ Earth-Life Science Institute, Tokyo Institute of Technology\\ Tokyo 152-8550, Japan\\[0,2cm]
 $~^2$ Department of Mathematics and Computer Science\\ University of Southern Denmark, Denmark\\[0.2cm]
 $^{3}$ Research Network Chemistry Meets Microbiology\\University of Vienna,  Wien A-1090, Austria \\[0.2cm]
$~^4$ Institute for Theoretical Chemistry\\ University of Vienna, Austria\\[0.2cm]
$~^5$ Department of Computer Science\\ University of Leipzig, Germany\\[0.2cm]
$~^6$ Max Planck Institute for Mathematics in the Sciences\\ Leipzig, Germany\\[0.2cm]
$~^7$ Fraunhofer Institute for Cell Therapy and Immunology\\ Leipzig, Germany\\[0.2cm]
$~^8$ Center for non-coding RNA in Technology and Health\\  University of Copenhagen, Denmark\\[0.2cm]
$~^9$ Santa Fe Institute, Santa Fe, USA
}
\date{}





\maketitle

\begin{abstract}
Computational techniques are required for narrowing down the vast space of
possibilities to plausible prebiotic scenarios, since precise information
on the molecular composition, the dominant reaction chemistry, and the
conditions for that era are scarce. The exploration of large chemical
reaction networks is a central aspect in this endeavour. While quantum
chemical methods can accurately predict the structures and reactivities of
small molecules, they are not efficient enough to cope with large-scale
reaction systems. The formalization of chemical reactions as graph grammars
provides a generative system, well grounded in category theory, at the
right level of abstraction for the analysis of large and complex reaction
networks. An extension of the basic formalism into the realm of integer
hyperflows allows for the identification of complex reaction patterns,
such as auto-catalysis, in large reaction networks using optimization
techniques.

\end{abstract}


\section{Introduction}

From a fundamental physics point of view chemical systems, or more
precisely molecules and their reactions, are just time dependent
multi-particle quantum systems, completely described by the fundamental
principles of quantum field theory (QFT) \cite{Weinberg:05}.
At this level of description
almost all questions of interest to a chemist are not tractable in
practice, however. A hierarchy of approximations and simplifications
is employed therefore to reach models of more practical value. These
are guided at least in part by conceptual notions that distinguish
chemistry from other quantum systems. Among these are constraints such
as the immutability of atomic nuclei and the idea that chemical
reactions comprise only a redistribution of electrons. On the formal
side the Born-Oppenheimer approximation \cite{Born:27} stipulates a
complete separation of the wave function of nuclei and electrons and
leads to the concept of the potential energy surface (PES) that
explains molecular geometries and provides a consistent -- if not
completely accurate -- view of chemical reactions as classical paths
of nuclear coordinates on the PES.  The PES itself is the result of
solving the Schr{\"o}dinger Equation with nuclear coordinates and
charges as parameters \cite{Mezey:87,Heidrich:91}. Quantum chemistry
(QC) has developed a plethora of computational schemes for this
purpose, typically trading off accuracy for computational resource
consumption. Among them in particular are the so-called semi-empirical
methods that use the fact the chemical bonds are usually formed by
pairs of electrons to decompose the electronic wave function into
contributions of electron pairs.

Molecular modelling (MM) and molecular dynamics (MD)
\cite{McCammon:77,Burkert:82} abandon quantum mechanics altogether and
instead treat chemical bond akin to classical springs. Sacrificing
accuracy, MM and MD can treat macro-molecules and supra-molecular complexes
that are well outside the reach of exact and even semi-empirical
quantum-chemical methods. For special classes of molecules, even coarser
approximations have been developed. Many properties of aromatic ring
systems, for instance, can be explained in terms of graph-theoretical
models known as H{\"u}ckel theory \cite{Hueckel:31,Hoffmann:63}. For
nucleic acids, on the other hand, models have been developed that aggregate
molecular building blocks (nucleotides) into elementary objects so that
Watson-Crick base pairs become edges in the graph representation
\cite{Zuker:81}.

A common theme in the construction of coarser approximations is that more
and more external information needs to be supplied to the model. While QFT
does not require more than a few fundamental constants, nuclear masses and
charges are given in practical QC computations. Semi-empirical methods,
in addition, require empirically determined parameters for electron correlation
effects. MM and MD models use extensive tables of parameters that specify
properties of localised bonds as a function of bond type and incident
atoms. Similarly, RNA folding depends on a plethora of empirically
determined energy contributions for base pair stacking and loop regions
\cite{Turner:10}.  A second feature of coarse-grained methods is that they
are specialised to answer different types of questions, or that different
classes of systems resort to different, mutually incompatible
approximations.

This well-established hierarchy of internally consistent models of
molecular structures is in stark contrast to our present capability to
model chemical reactions. While transition state theory \cite{Eyring:35}
does provide a means to infer reaction rate constants from PES, it requires
the prior knowledge of the educt and product states.

A systematic investigation of large chemical reaction systems requires the
development of a theoretical framework that is sufficiently coarse-grained
to be computationally tractable. Any such model must satisfy consistency
conditions that are inherited from the underlying physics.
\begin{figure}[!h]
  \includegraphics[width=\textwidth]{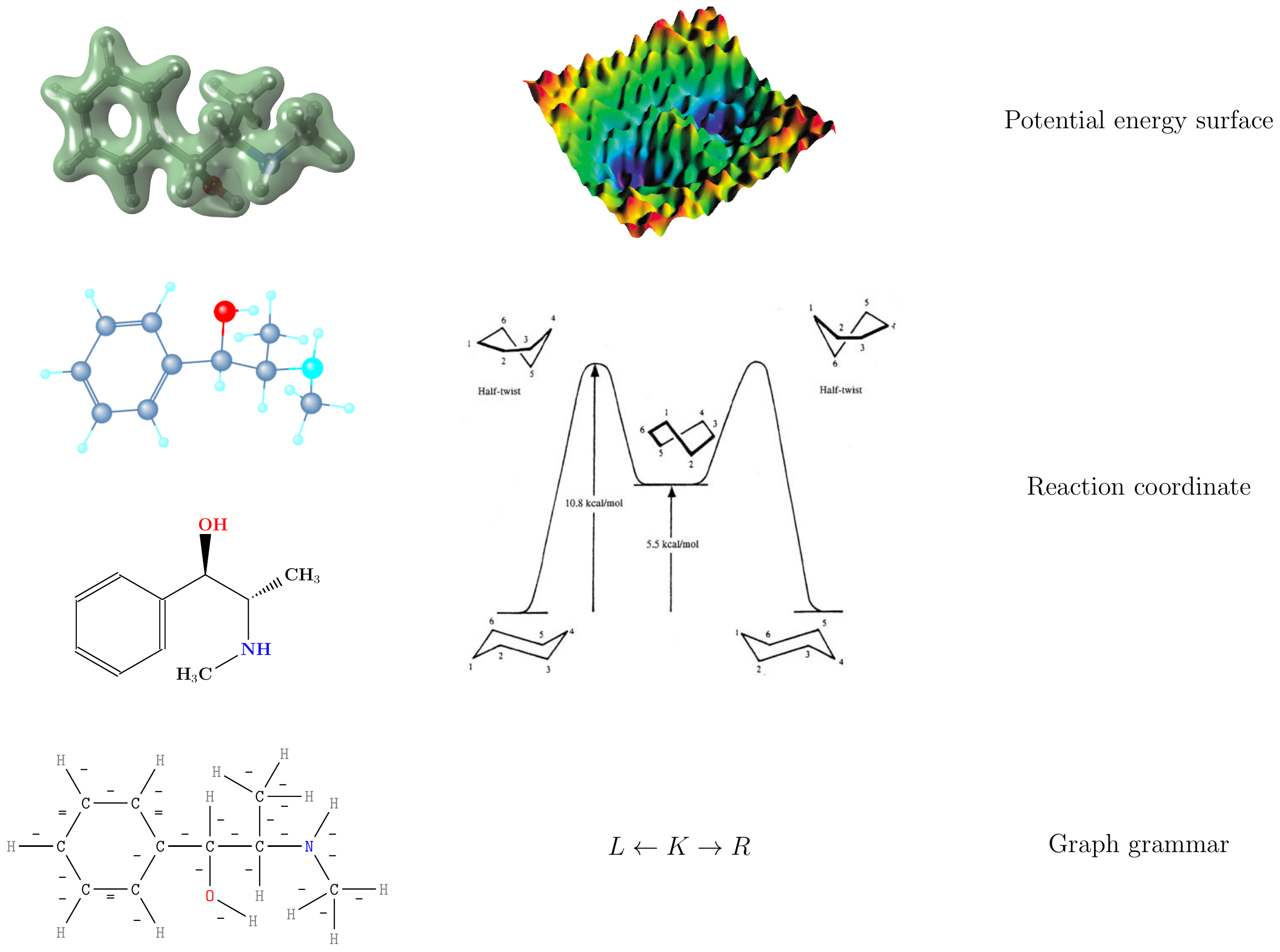}
  \caption{Levels of abstraction for computational approaches in
    chemistry. Shown is the hierarchy of approximations from quantum
    mechanics on the top to graph grammars on the bottom. The
    coarse-graining via the introduction of constraints (such as the
    Born-Oppenheimer approximation, or reducing coordinates of spatial
    objects to neighbourhood relations on graphs) is accompanied by a
    dramatic speedup in computation time.}
  \label{fig:levels}
\end{figure}
A recent study \cite{Rappoport:2014} on complex chemical reaction network
supports this view and concludes, that on the structure and reactivity
level of small molecules efficent QM based computational approaches exist,
but on the large-scale network level heuristic approaches are
indispensable. Here we argue that chemistry offers a coarse-grained level
of description that allows the construction of mathematically sound and
consistent formal models that, nevertheless, are conceptually and
structurally different from the formalism of quantum chemistry. Much of
chemistry is taught in terms of abstracted molecular structures and rules
(named reactions) that transform molecular graphs into each other. In the
following section we show how this level of modelling can indeed be made
mathematically precise and how it can accommodate key concepts of chemistry
such as transition state theory and fundamental conservation laws inherited
from the underlying physics.

\section{Graph Grammar Chemistry} 

The starting point of an inherently discrete model of chemistry is a
simplified, graphical representation of molecules. This coarse-graining
maps the atoms and bonds of a molecule to vertices and edges of the
corresponding graph. All type information (atom types C, H, N, O, etc.\ and
bond multiplicity single, double, triple bond) are mapped to labels on the
respective vertices or edges of the graph. In this setting all information
on properties which are tied to the three-dimensional space, such as
chirality, and cis/trans isomerism of double bonds is neglected. 

It is possible, however, to extend the model to retain the local geometric
information and thus capture the part
  of stereochemistry which is tied to stereogenic centers. Helicity,
  for instance, can not be expressed by the extended model since this
  property is generated by an extended spacial arrangement around an axis
  and not a single point or center. 
  The basis for the extended model is the valence-shell
electron-pair repulsion (VSEPR) model, which albeit comprising a set of
simple rules, has a firm grounding in quantum chemical modelling
\cite{Gillespie:08}. VSEPR theory determines approximate bond angles around
an atom depending on the incident bond types, i.e., in terms of information
conveyed by the labelled graph representation.  Stereochemical information
involving chiral centres as well as cis/trans isomerism thus can be encoded
simply in the order in which bonds are listed, and augmenting the labelled
graph with local permutation groups.

Still certain aspects of chemistry cannot be described in this form. For
instance, the concept of bonds as edges fails in multi-centre bonds since 
three or more atoms share a pair of bonding electrons. These are frequently
observed in boranes or organometallic compounds such as ferrocenes.
Non-local chirality, found for instance in helical molecules does not rely
on local, atom-centred symmetries and thus is not captured by local
orientation information. 

With molecules represented as graphs, the mechanism of a chemical reaction
is naturally expressed as a graph transformation rule. Graph transformation
thus retains the semantics familiar from organic chemistry textbooks.  As a
research discipline in computer science, graph transformation dates back to
the 1970s. Graph transformation has been studied extensively in the context
of formal language theory, pattern recognition, software engineering,
concurrency theory, compiler construction, verification among other fields
in computer science \cite{handbook1}.  Several formalisms have been
developed in order to formalize and implement the process of transforming
graphs.  Algebraic approaches are of particular interest for modelling
chemistry, where multiple variations based on category theory exist.
For example, different semantics can be expressed using either
the Single Pushout (SPO) approach, the more restrictive Double Pushout (DPO) approach,
or the recently developed Sesqui-Pushout (SqPO) approach \cite{Ses06}.

In the context of chemical reactions, DPO graph transformation is the
formalism of choice because it facilitates the construction of
transformation rules that are chemical in nature. DPO guarantees that all
chemical reactions are reversible \cite{Andersen:2013a}. The
conservation of atoms translates to a simple formal condition (formally,
the graph morphisms relating the context to the left hand and right hand
side of a rule must be bijections for vertices). In turn this requirement
guarantees the existence of well-defined atom maps.

\begin{figure}[!h]
  \includegraphics[width=\textwidth]{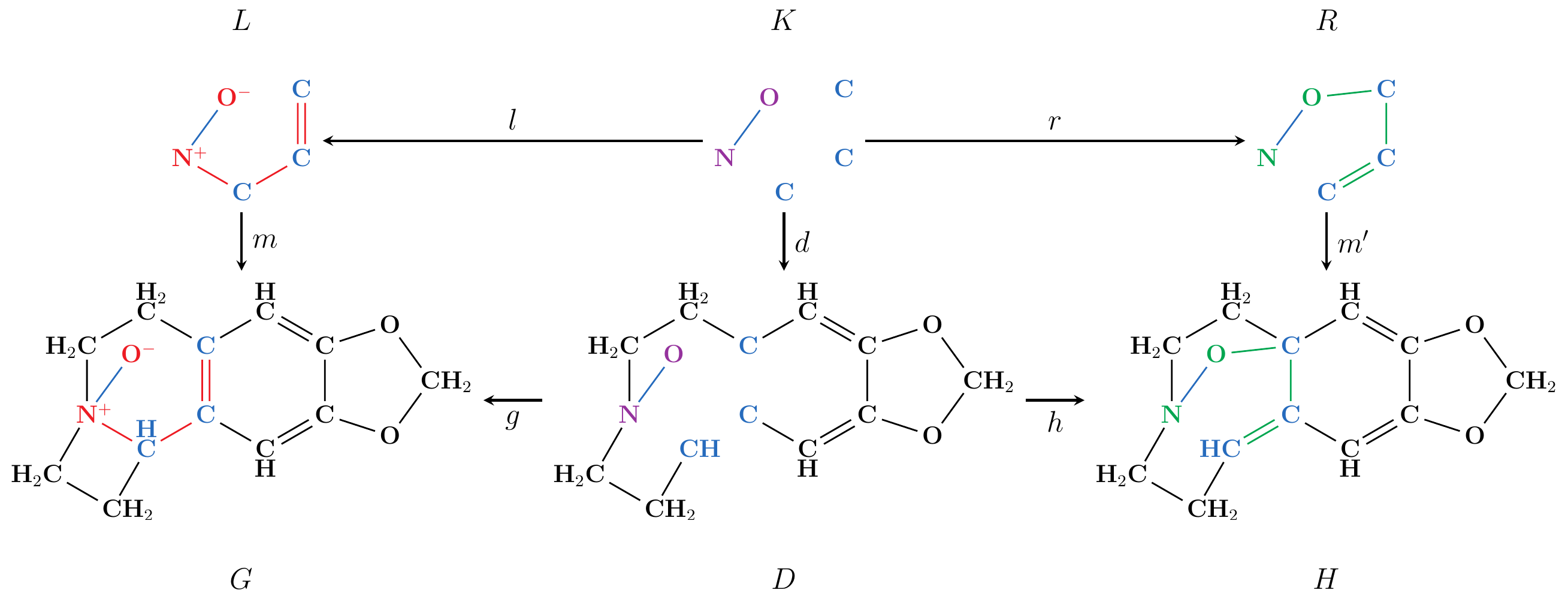}
  \caption{Double-pushout (DPO) representation of the application of a graph transformation rule. The
    actual reaction (the top span $L\leftarrow K\rightarrow R$) is the Meisenheimer rearrangement which transforms educt
    graph $G$ into product graph $H$. All arrows in the diagram are
    morphisms, i.e., functions which map vertices/edges from the graph on the
    arrow tail to the graph at arrow head. In order to be a valid transformation
    the two squares of the diagram must form so-called \emph{pushouts}.}
  \label{fig:DPO}
\end{figure}

Figure~\ref{fig:DPO}
illustrates how the Meisenheimer rearrangement \cite{Meisenheimer:1919}, a
temperature induced rearrangement of aliphatic amine oxides into
N-alkoxylamines, translates to the DPO formalism. The reaction transforms
educt graph $G$ (amine oxide) into product graph $H$ (N-alcoxylamine).
All arrows in the diagram are morphisms. The reaction centre, i.e., the subset of
atoms and bonds of the reactant molecules directly involved in the bond
breaking/forming steps of the chemical reaction, is expressed as a graph
transformation rule. The information of how to change the connectivity and
the charges of the atoms is specified by three graphs $(L, K, R)$. The left
graph $L$ (resp.\ right graph $R$) expresses the local state of molecules
before (resp.\ after) applying the reaction rule.  The context graph $K$
encodes the invariant part of the reaction centre and mathematical relates
$L$ and $R$ to each other.  The left graph $L$ is thus the precondition for
application of the rule (i.e., it can only be applied if there exists a
subgraph match $m$ that embeds $L$ in the host graph $G$; see the red and blue
highlighted part of graph $G$ in Figure~\ref{fig:DPO}). In this case $L$
can be replaced by the right graph $R$ (see the green and blue highlighted part of
graph $H$ in Figure~\ref{fig:DPO}), which transforms the educt graph $G$ into
the product graph $H$.

A computationally very demanding step when performing graph
transformation (e.g., for generating large chemical spaces) is the
enumeration of subgraph matches. Deciding if a single subgraph match
exists in a host graph is known to be an $\mathcal{NP}$-complete problem
\cite{Cook1971,Garey:79}.  Better theoretical results exists for certain
classes of graphs, e.g., for the so-called partial $k$-trees of bounded
degree (to which almost all molecule graphs belong
\cite{Yamaguchi:2004,Horvath:2010,Akutsua:2013}) where the subgraph
matching problem can be solved in polynomial time
\cite{Matousek:1992,Dessmark:2000}.  In practice it is however faster to
use simpler algorithms, e.g., VF2 \cite{vf2:1,vf2:2}.

\begin{figure}
\centering%
\includegraphics[width=9cm]{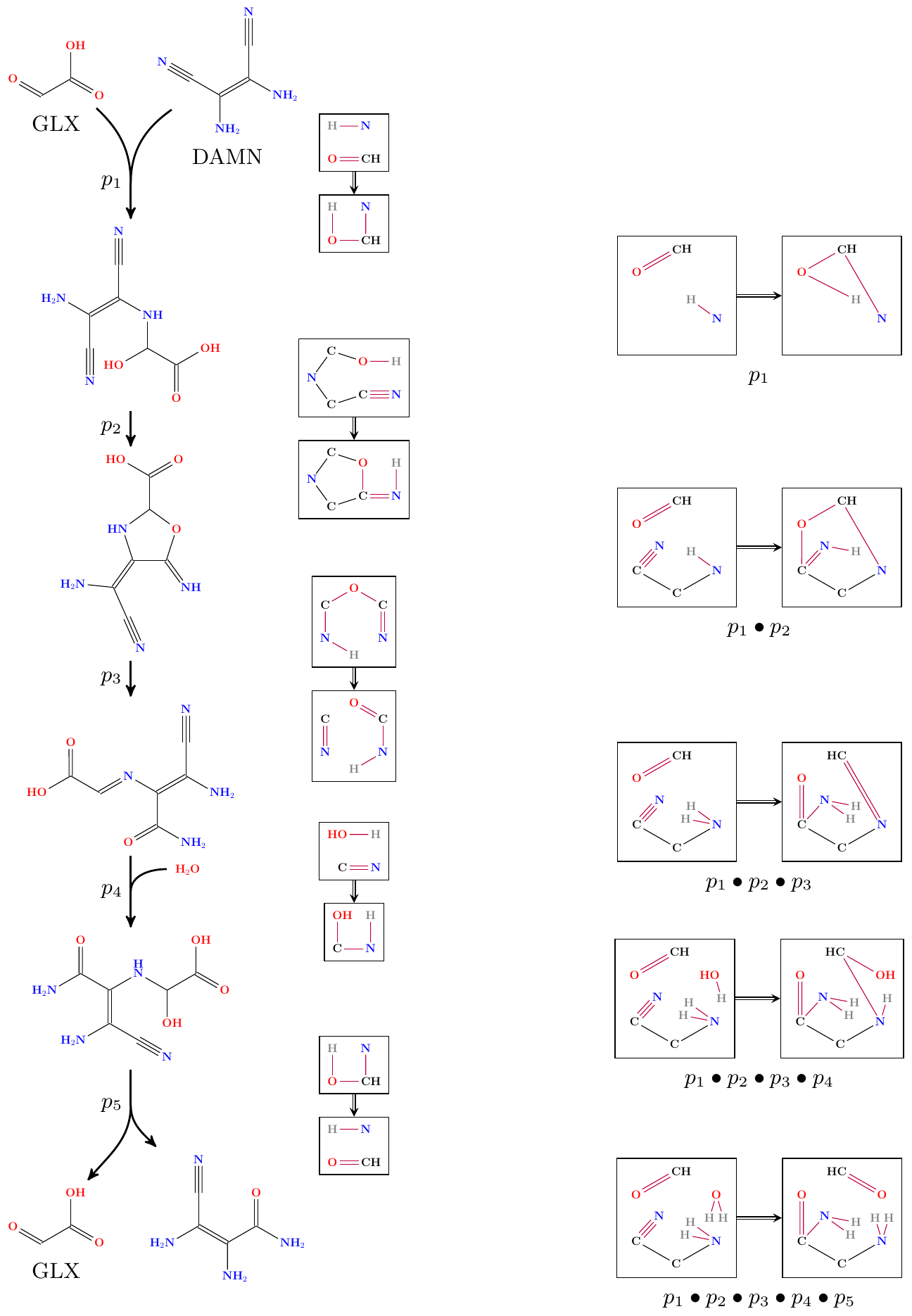}%
\caption[]{%
  Automatic inference of an overall rule by subsequent composition of graph transformation rules.
  The example is based on a sequence of reactions from
  \cite{esch07}, in which Eschenmoser describes how aldehydes act as
  catalysts for the hydrolysis of \chemfig{CN} groups of the
  \chemfig{HCN}-tetramer. The left column depicts the mechanism as
  presented in \cite{esch07}. An automated approach will, based on graph
  transformation rules, first generate a (potentially very large) chemical space
  (not depicted here). The depicted mechanism is then found as one of the
  solutions for the general question of enumerating hydrolysis pathways of
  \chemfig{HCN}-polymers that use glyoxylate (GLX: glyoxylate, CID~760) as
  catalyst. In the depicted pathway the tetramer of \chemfig{HCN} (DAMN:
  tetramer of \chemfig{HCN}, CID~2723951) is hydrolysed. The middle column
  depicts the left and right graph of each transformation rule $p_1,\ldots,p_5$
  that models the generalized reactions.
  The subsequently inferred rules $p_1$ (top), $p_1 \bullet p_2$, $\ldots$, and
  the overall rule $p_1 \bullet p_2 \bullet \ldots \bullet p_5$ (bottom)
  are depicted in the right column. Note, that there in general can be several
  composed overall rules, each expressing different atom traces.
  In this example there is only a single one. }
\label{fig:overall}
\end{figure}

Chemical reactions are often compositions of elementary reactions. In the
latter, the reaction centre can always be expressed as a
cycle\cite{Hendrickson:1997,Hendrickson:2010}, with an even number of
vertices for homovalent reactions and an odd number from ambivalent
reactions, better known as redox reactions. Graph transformations have a
natural mechanism for rule composition that allows the expression of
multi-step reactions (e.g., enzyme-mediated reactions or even complete metabolic
pathways) as compositions of elementary transformation rules. The
properties of elementary rules in terms of mass conservation or
atom-to-atom mapping nicely carry over to the composed ``overall
transformation rules''.  Since the action of chemical reactions is to
redistribute atoms along complex reaction sequences, rule composition can
be used to study the trace of individual atoms along these reaction
sequences in a chemically as well as mathematically correct fashion. Rule
composition can be completely automatized and thus opens the possibility
for model reduction (see \cite{Andersen:2014c} for further details). We
illustrate rule composition in the context of prebiotic chemistry in
Figure~\ref{fig:overall}.

\section{Network View of Chemistry}

Classical synthetic chemistry traditionally has been concerned with the
step-wise application of chemical reactions in carefully crafted synthesis
plans. In living organisms, in contrast, complex networks of intertwined
reactions are active concurrently. These intricate reaction webs harbour
complex reaction patterns such as branch points, autocatalytic cycles, and
interferences between reaction sequences. The emerging field of Systems
Chemistry has set out to leverage the systemic, network-centred view as a
framework also for synthetic chemistry. Consequently, large-scale chemical
networks are no longer just a subject of analysis in the context of
understanding the working of a living cell's metabolism, but are becoming a
prerequisite to understanding the possibilities within chemical
spaces, i.e., the universe of chemical compounds and the possible
  chemical reactions connecting them.
 The formulation of a predictive theory of chemical space
requires it to be rooted in a strict mathematical formalization and
abstraction of the overwhelming amount of anecdotal knowledge, that has
been collected on the single reaction and functional subnetwork level, into
generalizing principles.

Graph transformation systems, as discussed earlier, provide the basis for
such a formalism that allow for a systematic and step-wise construction of
arbitrary chemical spaces.  A chemical system is then specified as a formal
graph grammar that encapsulate a set of transformation rules, encoding the
reaction chemistry, together with a set of molecules which provide the
starting points for rule application.  The iteration of the graph grammar
yields reaction networks in the form of directed hypergraphs as explicit
instantiations of the chemical space. Usually a simple iterative expansion
of the chemical space leads to a combinatorial explosion in the number of
novel molecules. Therefore a sophisticated strategy framework for the
targeted exploration of the parts of interest of the chemical space has
been developed \cite{Andersen:2014a}. Such strategies are indispensable if
for example polymerization/cyclization spaces are the subject of
investigation.  These type of spaces are for example found in the important
natural product classes of polyketides and terpenes, and in prebiotic
\chemfig{HCN} chemistry \cite{Andersen:2013b}.  The strategy framework
allows the guidance of chemical space exploration not only using
physico-chemical properties of the generated molecules, but also using
experimental data such as mass spectra.  Importantly, the hypergraphs
(reaction networks) are generated automatically annotated with atom-to-atom
maps, as defined implicitly in the underlying graph grammar.  For large and
complex reaction networks it is thus possible to construct atom flow
networks in an automated fashion, even including corrections for molecule
and subnetwork symmetries, as required for the interpretation of isotope
labelling experiments \cite{Buescher:2015}.

The origin of life can be viewed as an intricate process which has been
shaped by external constraints provided by early Earth's environment and
intrinsic constraints stemming from reaction chemistry itself. Higher order
chemical transformation motifs, such as network auto-catalysis, are
believed to have played a key role in the amplification of the building
blocks of life \cite{Kauffman:95, Kauffman:07, Ulanowicz:97}.
A combination of the constructive graph grammar approach
with techniques from combinatorial optimization sets the proper formal
stage for attacking some of these origin of life related questions.  The
key idea here is to rephrase the topological requirements for a particular
chemical behaviour, e.g., network auto-catalysis, as an optimization problem
on the underlying reaction network (hypergraph).  An example of this is the
enumeration of pathways with specific properties, which can be formally
modelled as a constrained hyperflow problem.  Many of
these problems are theoretically computationally hard \cite{andersen:12},
though in practice methods such as Integer Linear Programming (ILP) can be
successfully used to identify such transformation motifs in arbitrary chemical
spaces.  The enumeration of transformation motifs is the first step in
computer-assisted large-scale analysis of reaction networks. Other
mathematical formalisms, such as Petri nets, and in general concurrency
theory, can subsequently be used to model properties of chemical systems on
an even higher-level.  Complicated chemical spaces, such as the one formed by the
formose process \cite{Butlerov:1861}, can thus be dismantled into coupled
functional modules, advancing the understanding of how a particular
reaction chemistry induces specified behaviour on the reaction network
level. More generally speaking and emphasizing the need for new
  approaches, it is well foreseeable that the future of chemistry is
  strongly bundled with a deeper understanding of complex chemical
  systems \cite{Whitesides:15,Rewire}, and the necessary skills to
  analyze such systems will become more and more important.









\section{Discussion and Concluding Remarks} 
Narrowing down potential pathways for prebiotic scenarios
indispensably requires novel systemic approaches that allow for the
investigation of large chemical reaction systems. While the
development of mathematically well grounded methods for abstraction
and coarse-graining of (concurrent) systems is a very active research
area in computer science, the interdisciplinary endeavor to integrate
these approaches with chemistry is more often treated as a conceptual
possibility rather than as a predictive approach. Many of the problems
to be solved in this process are computationally hard (i.e.,
$\mathcal{NP}$-hard) but still allow for practical {\em in silico}
solutions. This discrepancy led to a relatively new and successful
subfield in computer science called algorithmic engineering
\cite{Sanders09}, in which one of the goals is to bridge the gap
between theoretical results and practical solutions to hard
problems. Clearly, results from that field should be taken into
account when large chemical systems with a plethora of underlying hard
problems have to be solved.
%
%
%

As an illustration of the integrative potential we sketch an
example in Figure~\ref{fig:overall} (see \url{http://mod.imada.sdu.dk}
for further examples). It shows how graph transformation based
chemical space exploration (rooted in graph theory, category theory,
and concurrency theory) with subsequent solution enumeration (using
diverse optimization techniques) can be applied to a reaction schema
presented in \cite{esch07}, in which Eschenmoser describes how
aldehydes act as catalysts for the hydrolysis of \chemfig{CN} groups
of the \chemfig{HCN}-tetramer. Given a set of chemical reactions
$p_1,\ldots,p_5$ (encoded as graph grammar rules) and a set of initial
molecules, the iterative application of these rules (potentially with
an underlying strategy for the space expansion) leads to a chemical
space encoded as hypergraph. This hypergraph is the source for
solving the subsequent problem of inferring and enumerating
declaratively defined reaction motifs or pathways. In
Figure~\ref{fig:overall} we do not illustrate the expansion step, the
depicted mechanism (left column) is however found automatically as one
of the potentially many solutions for the general question of
enumerating hydrolysis pathways of \chemfig{HCN}-polymers that use
glyoxylate (GLX: glyoxylate, CID~760) as catalyst. Formally such a
solution is encoded as an integer hyperflow within the underlying
hypergraph. Given the depicted reaction sequence, the possibility for
transformation rule composition is utilised. The overall, (more
coarse-grained) rule $p_1 \bullet p_2 \bullet \ldots \bullet p_5$ is thus
automatically inferred by consecutive composition of the
(simpler) transformation rules $p_1, \ldots, p_5$. All intermediate steps are
depicted in the right column of Figure~\ref{fig:overall}. Note, that
the automated coarse-graining implemented by rule composition allows
for keeping track of the possibilities of different atom traces, expressed as the
atom-atom mapping from educt to product in composed rules. An obvious
reachable next step is therefore the analysis as well as the design of
isotope labeling experimentation based on the \textit{in silico}
generative chemistry approach with subsequent trace analysis.
%
%

Clearly, the illustration in Figure~\ref{fig:overall} is serving
only as  an example. The modelling of essential chemical
parameters including kinetic components and thermodynamics are still
missing. Nevertheless, the approach is already highly automated, and
will bring wetlab and \textit{in silico} experiments closer
together. We argue that the intermediate-level theory outlined here
holds promise in many fields of chemistry. In particular, we
suggest that it is a plausible substrate for a predictive theory of
prebiotic chemistry.

\section*{Acknowledgements}
This work is supported by the Danish Council for Independent
  Research, Natural Sciences, the COST Action CM1304 ``Emergence and
  Evolution of Complex Chemical Systems'', and the ELSI Origins Network
  (EON), which is supported by a grant from the John Templeton
  Foundation.The opinions expressed in this publication are those of the
  authors and do not necessarily reflect the views of the John Templeton
  Foundation.


\begingroup
\newcommand\doi[1]{(\href{http://dx.doi.org/#1}{doi:#1})}
\renewcommand{\providecommand}[2]{}{}
\bibliography{arxiv}
\endgroup

\end{document}